\documentclass[preprint,12pt]{article}
\usepackage{geometry}
\usepackage{amsmath, amssymb, amsthm}
\usepackage{authblk} 
\usepackage{indentfirst}
\usepackage{hyperref}
\usepackage{cite}

\usepackage{titlesec} 
\titleformat{\section}
{\normalfont\bfseries}                 
{\thesection.}                
{0.5em}                      
{}                           

\numberwithin{equation}{section} 

\title{(3+1)-dimensional modified Kadomtsev-Petviashvili equation and its $\bar{\partial}$-formalism}
\author[]{Yue Li}
\author[]{Fei Li}
\author[]{Mengli Tian}
\author[]{Yuqin Yao\thanks{Corresponding author.\\      \hangindent=1.85em\hangafter=1
		   E-mail address: yaoyq@cau.edu.cn (Y.Yao)}}
\affil[]{College of Science, China Agricultural University, Beijing, 100048, PR China}
\date{}

\begin{document}
	
	\maketitle
	
	\begin{abstract}
		
	Constructing integrable evolution nonlinear PDEs in three spatial dimensions is one of the most important open problems in the area of integrability. Fokas achieved progress in 2006 by constructing integrable nonlinear equations in 4+2 dimensions, but the reduction to 3+1 dimensions remained open until 2022 when he introduced a suitable nonlinear Fourier transform to achieve this reduction for the Kadomtsev-Petiashvili (KP) equation. Here, the integrable generalization of the modified KP (mKP) equation has been presented which has the novelty that it involves complex time and preserves Laplace's equation. The (3+1)-dimensional mKP equation is obtained by imposing the requirement of real time. Then, the spectral analysis of the eigenvalue equation is given and the solution of the $\bar{\partial}$-problem is demonstrated based on Cauchy-Green formula. Finally, a novel $\bar{\partial}$-formalism for the initial value problem of the (3+1)-dimensional mKP equation is worked out.
	
	\end{abstract}
	

	\section{Introduction}
	
	 The higher-dimensional generalizations of integrable systems hold profound significance in both physics and mathematics \cite{ref1}. In the realm of physics,  higher-dimensional integrable systems can more accurately describe multi-body interactions, physical phenomena in higher-dimensional spacetimes (such as the extra dimensions in string theory), or complex phase transitions in condensed matter systems. The study of higher-dimensional integrable systems not only deepens our understanding of nonlinear dynamics but also further solidifies the bridge between mathematics and physics. In the 1980s, Zakharov, Manakov, and Segur were among the first to study (2+1)-dimensional equations \cite{ref2}. In the 1990s, Fokas and Zakharov utilized the nonlocal Riemann-Hilbert method and $\bar{\partial}$-method to elevate the dimension from one to two \cite{ref3,ref4,ref5,ref6, ref7}. The question of constructing integrable evolution nonlinear PDEs in three spatial dimensions was one of the most important open problems in the area of integrability. Progress in this direction was achieved in 2006 and 2007 by Fokas who constructed integrable nonlinear equation in 4+2, i.e. in four spatial and two temporal dimensions \cite{ref8,ref9}. Although the reduction of such equations to 3+2, is straightforward, the reduction to 3+1 remained open until 2022, when Fokas introducing a suitable nonlinear Fourier transform achieved this reduction for the Kadomtsev-Petiashvili (KP) equation \cite{ref10}. In 2023, the Wronskian and Grammian determinant solutions of the (3+1)-dimensional KP equation are studied \cite{ref11}.
	
	$\bar{\partial}$-dressing method, an extension of the RH method, was introduced into the field through the research efforts by Beals and Coifman \cite{ref12}. In 2008, Fokas solved Cauchy problem for an integrable generalization of the KP equation in 4+2 by $\bar{\partial}$-dressing method \cite{ref13}. 
	Consequently, in subsequent studies of higher-dimensional equations, $\bar{\partial}$-dressing method has played a significant role in finding various solutions through the Fourier transform of the Green's function \cite{ref10}. For example, it can be used to construct the solutions particularly of the (2+1)-dimensional Jimbo-Miwa equation and to seek for the line-soliton and rational solutions to the (2+1)-dimensional Boussinesq equation \cite{ref14,ref15}.
	
	The modified KP(mKP) equation is of the form
	\begin{equation}\label{eq:1.1}
		q_t + q_{xxx} - 6 q^2 q_x + 3 \partial_x^{-1} q_{yy} + 6 q_x \partial_x^{-1} q_y = 0, 
	\end{equation}
	which is a two-dimensional extension of the modified Korteweg-de Vries (mKdV) equation \cite{ref16}.
	The following linear problems can serve as the compatibility conditions for equation (1.1) \cite{ref17}
	\begin{equation}\label{eq:1.2}
		\Psi_y - \Psi_{xx} - 2 q \Psi_x = 0, 
	\end{equation}
	\begin{equation}\label{eq:1.3}
		\Psi_t + 4 \Psi_{xxx} + 12 q \Psi_{xx} + 6 \left(q_x + \partial_x^{-1} q_y + q^2\right) \Psi_x = 0, 
	\end{equation}
	which are exactly what is referred to as so-called the Lax pair.
	In our paper, we want to construct the integrable (3+1)-dimensional mKP equation and its Lax pair. It is interesting that it involves complex time and preserves Laplace's equation \cite{ref18}. Further, the $\bar{\partial}$-formulation to the (3+1)-dimensional mKP equation is proposed by using  $\bar{\partial}$-dressing method. 
	
	The paper is organized as follows: in Sec. 2, we present the complexification of the mKP equation to 4+2 and obtain the (3+1)-dimensional mKP equation. In Sec. 3, spectral analysis is conducted on the spatial part of the Lax pair, and the Fourier transform pair is introduced through the Green's function. In Sec. 4, the $\bar{\partial}$-formalism to the (3+1)-dimensional mKP equation is obtained. A conclusion is given in the last section.
	
	\section{Three Dimensional Extensions of the modified Kadomstsev-Petviashvili equation}
	
	For the convenience of the following analysis, we make a transformation
	$$
	\Psi = e^{- \partial_x^{-1} q} \Phi,
	$$
	then the Lax pair (1.2) and (1.3) can be transformed to 
	\begin{equation}\label{eq:2.1}
		\Phi_y - \Phi_{xx} +\left(q_x +q^2 - \partial_x^{-1} q_y\right) \Phi = 0, 
	\end{equation}
	\begin{equation}\label{eq:2.2}
		\Phi_t - \left(\partial_x^{-1} q_t +4 q_{xx} +6 q q_x - 2 q^3 + 6 \partial_x^{-1} q_y q \right) \Phi -\left(6 q_x + q^2 - 6 \partial_x^{-1} q_y\right) \Phi_x + 4 \Phi_{xxx}  = 0. 
	\end{equation}
	
	Supposing that the potential function $q(x, y)$ rapidly
	tends to zero when $(x,y) \rightarrow \pm \infty$, an asymptotic solution can be derived as
	$$
	\Phi \thicksim e^{i k x - k^2 y} , (x,y) \rightarrow \pm \infty,
	$$
	which is called the Jost solution and parameter $k$ is an arbitrary. Further, making a transformation
	\begin{equation}\label{eq:2.3}
		\mu(x,y,k) = \Phi(x,y,k) e^{- i k x + k^2 y},
	\end{equation}
	with $\mu(x,y,k) \thicksim I , (x,y) \rightarrow \pm \infty $, the Lax pair (2.1) and (2.2) can be written as
	\begin{equation}\label{eq:2.4}
		\mu_y - \mu_{xx} -2 i k \mu_x + \left(q_x + q^2 - \partial_x^{-1} q_y\right) \mu = 0,
	\end{equation}
	\begin{equation}\label{eq:2.5}
		\begin{aligned}
			&
			\mu_t - \left(\partial_x^{-1} q_t + 4 q_{xx} + 4 i k^3 - 2 q^3 + 6 i k q_x +6 q q_x + 6 i k q^2 + 6 \partial_x^{-1} q_y q - 6 i k \partial_x^{-1} q_y \right) \mu
			\\
			&
			- \left(12 k^3 + 6 q_x + 6 q^2 - 6 \partial_x^{-1} q_y\right) \mu_x -12 i k \mu_{xx} + 4 \mu_{xxx} = 0.
		\end{aligned}
	\end{equation}
	
	Starting from the (2+1)-dimensional mKP equation, it has been demonstrated that a (4+2)-dimensional equation is achieved by complexifying the independent variables $x,  y, t$ as follows:
	\begin{equation}\label{eq:2.6}
		q_{\bar{t}} + q_{\bar{x}\bar{x}\bar{x}} - 6 q^2 q_{\bar{x}} + 3 \partial_{\bar{x}}^{-1} q_{\bar{y}\bar{y}} + 6 q_{\bar{x}} \partial_{\bar{x}}^{-1} q_{\bar{y}} = 0,
	\end{equation}
	and its Lax pair becomes
	\begin{equation}\label{eq:2.7}
		\mu_{\bar{y}} - \mu_{\bar{x}\bar{x}} -2 i k \mu_{\bar{x}} + \left(q_{\bar{x}} + q^2 - \partial_{\bar{x}}^{-1} q_{\bar{y}}\right) \mu = 0,
	\end{equation}
	\begin{equation}\label{eq:2.8}
		\begin{aligned}
			&
			\mu_{\bar{t}} - \left(\partial_{\bar{x}}^{-1} q_{\bar{t}} + 4 q_{\bar{x}\bar{x}} + 4 i k^3 - 2 q^3 + 6 i k q_{\bar{x}} +6 q q_{\bar{x}} + 6 i k q^2 + 6 \partial_{\bar{x}}^{-1} q_y q - 6 i k \partial_{\bar{x}}^{-1} q_{\bar{y}} \right) \mu 
			\\
			&
			- \left(12 k^3 + 6 q_{\bar{x}} + 6 q^2 - 6 \partial_{\bar{x}}^{-1} q_{\bar{y}}\right) \mu_{\bar{x}} -12 i k \mu_{\bar{x}\bar{x}} + 4 \mu_{\bar{x}\bar{x}\bar{x}} = 0,
		\end{aligned}
	\end{equation}
	where 
	\begin{equation}\label{eq:2.9}
		\begin{aligned}
		&t = t_1 + it_2, x = x_1 + ix_2, y = y_1 + iy_2, 
		\\
		&
		t_1, t_2, x_1, x_2, y_1, y_2 \in \mathbb{R}.
		\end{aligned}
	\end{equation}
	
	 The variable $q(x_1, x_2, y_1, y_2, t_1, t_2)$ depends on four spatial variables $x_1, x_2, y_1, y_2$ and two temporal variables $t_1, t_2$.
	Based on equation (2.9), the following relationships hold
	\begin{equation}\label{eq:2.10}
		\partial_{\bar{t}} = \frac{1}{2}(\partial_{t_1} + i\partial_{t_2}), \\
		\partial_{\bar{x}} = \frac{1}{2}(\partial_{x_1} + i\partial_{x_2}), \\
		\partial_{\bar{y}} = \frac{1}{2}(\partial_{y_1} + i\partial_{y_2}),
	\end{equation}
	
	\begin{equation}\label{eq:2.11}
		(\partial_{\bar{x}}^{-1} f)(x_1, x_2) = \frac{1}{\pi} \iint_{\mathbb{R}^2} \frac{f(x_1', x_2')}{x - x'} \, dx_1' \, dx_2'.
	\end{equation}
	
	\noindent\textbf{Remark 2.1.} 
	\textit{Equation (2.6) admits the following interesting reductions for q is real and satisfying both} $q_{\bar{x}y}=0$ \textit{and the Laplace's equation, that is}
	\begin{equation}\label{eq:2.12}
		\begin{aligned}
		q_{t_1} = -q_{{x_1}{x_1}{x_1}} +& 6q^2q_{x_1} - 3\partial_{x_1}^{-1}q_{{y_1}{y_1}} - 6q_{x_1}\partial_{x_1}^{-1}q_{{y_1}},
		\\
		&
		q_{x_1 x_1} + q_{x_2 x_2}=0,
		\end{aligned}
	\end{equation}
	\textit{and}
	\begin{equation}\label{eq:2.13}
		\begin{aligned}
			q_{t_2} = q_{{x_2}{x_2}{x_2}} +& 6q^2q_{x_2} + 3\partial_{x_2}^{-1}q_{{y_2}{y_2}} - 6q_{x_2}\partial_{x_2}^{-1}q_{{y_2}},
			\\
			&
			q_{x_1 x_1} + q_{x_2 x_2}=0.
		\end{aligned}
	\end{equation}
	
	Now we formulate the relational equations
	\begin{equation}\label{eq:2.14}
		\xi = ax_1 + bx_2, \tau = \tilde{a}t_1 + \tilde{b}t_2,
	\end{equation}
	where $a, b, \tilde{a}, \tilde{b}$ are real constants, we obtain
	\begin{equation}\label{eq:2.15}
		\partial_{\bar{x}} = K \frac{\partial}{\partial\xi} , \partial_{\bar{t}} = \tilde{K} \frac{\partial}{\partial\tau},
	\end{equation}
	where
	$$
	\ K = \frac{a + ib}{2}, \tilde{K} = \frac{\tilde{a} + i\tilde{b}}{2}.
	$$
	
	Thus, by incorporating transformation (2.15), the  (4+2)-dimensional mKP equation (2.6) can be transformed to
	\begin{equation}\label{eq:2.16}
		\tilde{K} q_{\tau} + K^3 q_{\xi\xi\xi} - 6 K q^2 q_{\xi} + 3 K^{-1} \partial_{\xi}^{-1} q_{\bar{y}\bar{y}} + 6 q_\xi \partial_{\xi}^{-1} q_{\bar{y}}= 0.
	\end{equation}
	
	Further, performing the transformation
	\begin{equation}\label{eq:2.17}
		q \rightarrow Q q, \bar{y} \rightarrow \frac{\bar{y}}{Y}, k \rightarrow K k,
	\end{equation}
	 where $ Q $ and $ Y $ are complex constants, we can obtain
	\begin{equation}\label{eq:2.18}
		q_{\tau} + \alpha q_{\xi\xi\xi} - \beta q^2 q_{\xi} + \gamma \partial_{\xi}^{-1}  q_{\bar{y}\bar{y}} + \sqrt{2 \beta \gamma} q_\xi \partial_{\xi}^{-1} q_{\bar{y}}= 0,
	\end{equation}
	where
	\begin{equation}\label{eq:2.19}
		\alpha = \frac{K^3}{\tilde{K}},\beta = \frac{6 K Q^2}{\tilde{K}},\gamma = \frac{3 Y^2}{K \tilde{K}}.
	\end{equation}
	
	Applying transformations (2.15) and (2.17) on the Lax pair (2.7) and (2.8), 
	and renaming $\tau, \xi, y_1, y_2$ to $t, x, y, z$ and letting $\gamma$ to $4\gamma$ , we obtain the (3+1)-dimensional mKP equation
	\begin{equation}\label{eq:2.20}
		q_{t} + \alpha q_{xxx} - \beta q^2 q_x + \gamma \partial_x^{-1} \left(q_{yy}-q_{zz}+2iq_{yz}\right) + \sqrt{2 \beta \gamma} q_x \partial_x^{-1} \left(q_y+iq_z\right)= 0,
	\end{equation}
	and its Lax pair  
	\begin{equation}\label{eq:2.21}
		\begin{small}
		\left[ 
		\left(\frac{\gamma}{3\alpha} \right)^{1/2} \left(\partial_y + i\partial_z\right) - \partial_x^2 - 2ik  \partial_x + \left(\frac{\beta}{6\alpha} \right)^{1/2} q_x + \frac{\beta}{6\alpha}q^2 -\partial_x^{-1} \left(\frac{\beta}{6\alpha}  \right)^{1/2} \left(\frac{\gamma}{3\alpha} \right)^{1/2} \left(q_y + iq_z \right) 
		\right] \mu = 0,
		\end{small}
	\end{equation}
	\begin{equation}\label{eq:2.22}
		\begin{small}
		\begin{aligned}
			&
			\mu_t - \partial_x^{-1} \sqrt{\frac{\beta}{6\alpha}} q_t \mu - 4\alpha \left(\sqrt{\frac{\beta}{6\alpha}} q_{xx} - \partial_x^3 - 3ik\partial_x^2 +3k^2 \partial_x + ik^3\right) \mu + 2 \alpha \left(\frac{\beta}{6\alpha}\right)^{3/2} q^3 \mu -6\alpha
			\\
			&
			 \left[\sqrt{\frac{\beta}{6\alpha}} q_x  \left(\partial_x + ik\right) + \frac{\beta}{6\alpha}q q_x + \frac{\beta}{6\alpha} q^2 \left(\partial_x + ik\right)\right] \mu
			- 6\alpha \partial_x^{-1} \sqrt{\frac{\beta}{6\alpha}} \sqrt{\frac{\gamma}{3\alpha}} \left(q_y + iq_z \right) \left(\sqrt{\frac{\beta}{6\alpha}} q - \partial_x - ik\right)\mu = 0.
		\end{aligned}
    	\end{small}
	\end{equation}
	
	\section{Spectral Analysis}
	Focus on the spatial part of the Lax pair (2.21), if $\gamma/3\alpha$ is a real number, replacing $\left( \gamma/3\alpha \right)^{1/2} y$  with $y$, $\left( \gamma/3\alpha \right)^{1/2} z$ with $z$, and scaling $q$, it becomes
	\begin{equation}\label{eq:3.1}
		\left\{\partial_y + i\partial_z - \partial_x^2 - 2ik\partial_x + \left[q_x+q^2-\partial_x^{-1}\left(q_y+iq_z\right)\right] \right\} \mu(x,y,z,k) = 0,
	\end{equation}
	if $\gamma/3\alpha$ is a complex number, equation (3.1) can be directly obtained by scaling $y$ and $z$.
	
	\noindent\textbf{Proposition 3.1.} 
	\textit{Assuming $q(x,y,z) \in S(\mathbb{R}^3) $ denotes the space of Schwartz functions in three dimensions and vanishes at infinity, the solution of the eigenvalue equation (3.1) is obtained as follows:} 
	\begin{equation}\label{eq:3.2}
		\mu(x, y, z, k) = 1 - \frac{1}{\pi} \int_{\mathbb{R}^3} 
		\frac{\xi e^{i\left[\xi x - 2h_I \xi y + (\xi^2 + 2\xi h_R)z\right]}}{h - k} 
		\widehat{q}(\xi, h) 
		\mu(x, y, z, h + \xi) 
		\, d\xi \, dh_R \, dh_I,
	\end{equation}
	$$
	k \in \mathbb{C}, h=h_R+ih_I,
	$$
	\textit{where}
	\begin{equation}\label{eq:3.3}
		\widehat{q}(\xi, k) = -\frac{1}{4\pi^2} \int_{\mathbb{R}^3} 
		e^{-i\left[\xi x' - 2k_I \xi y' + (\xi^2 + 2\xi k_R) z'\right]}\left[\partial_x^{-1}\left(q_y+iq_z\right)-q_x-q^2\right]\mu(x',y',z',k)\,dx'dy'dz'.
	\end{equation}
	
	\noindent\textbf{Proof.}
	 By performing the spectral analysis, the solution of equation (3.1) satisfies the following linear integrable equation
	\begin{equation}\label{eq:3.4}
		\begin{aligned}
			\mu(x,y,z,k) &= 1 + \int_{\mathbb{R}^3} G(x-x',y-y',z-z',k)
			\\
			&
			\left[\partial_x^{-1}\left(q_y+iq_z\right)-q_x-q^2\right]\mu(x',y',z',k)dx'dy'dz',
		\end{aligned}
	\end{equation}
	with
	$$
	\mu(x,y,z,k) \thicksim 1, x,y,z \rightarrow \pm \infty.
	$$
	
	Apply the Fourier transform to the linear terms on both sides of equation (3.1), the Green's function is defined as follows: 
	\begin{equation}\label{eq:3.5}
		G_y+iG_z-G_{xx}-2ikG_x = \delta\left(x\right) \delta\left(y\right) \delta\left(z\right), 
	\end{equation}
	with the identity
	$$
	\delta\left(x\right) \delta\left(y\right) \delta\left(z\right) = \frac{1}{(2\pi)^3}\int_{\mathbb{R}^3}e^{i(\xi x + \eta y + \zeta z)}d\xi d\eta d\zeta,
	$$
	thus, the solution of $G$ in equation (3.5) is given by
	\begin{equation}\label{eq:3.6}
		G(x,y,z,k) = -\frac{1}{(2\pi)^3}\int_{\mathbb{R}^3}\frac{e^{i(\xi x + \eta y + \zeta z)}}{i\eta-\zeta+\xi^2+2k\xi}\,d\xi d\eta d\zeta.
	\end{equation}
	
	Computing the $\frac{\partial\mu}{\partial\bar{k}}$ derivative to equation (3.4), we have 
	\begin{equation}\label{eq:3.7}
		\begin{aligned}
			\frac{\partial\mu}{\partial\bar{k}}(x,y,z,k)&= \chi(x,y,z,k)+\int_{\mathbb{R}^3} G(x-x',y-y',z-z',k)
			\\
			&
			\left[\partial_x^{-1}\left(q_y+iq_z\right)-q_x-q^2\right]\frac{\partial\mu}{\partial\bar{k}}(x',y',z',k)\,dx'dy'dz',
		\end{aligned}
	\end{equation}
	where
	\begin{equation}\label{eq:3.8}
		\chi(x,y,z,k)=\int_{\mathbb{R}^3} \frac{\partial G}{\partial\bar{k}}(x-x',y-y',z-z',k)\left[\partial_x^{-1}\left(q_y+iq_z\right)-q_x-q^2\right]\mu(x',y',z',k)\,dx'dy'dz'.
	\end{equation}
	
	Here, an identity is mentioned, 
	\begin{equation}\label{eq:3.9}
		\frac{\partial}{\partial \bar{k}} \left( \frac{1}{k_1 - k_2} \right) 
		= \pi \delta(k_{1R} - k_{2R}) \delta(k_{1I} - k_{2I}).
	\end{equation}
	
	Now we calculate $\frac{\partial G}{\partial\bar{k}}$ as follows:
	\begin{equation}\label{eq:3.10}
	\frac{\partial G}{\partial\bar{k}}=-\frac{1}{(2\pi)^3}\int_{\mathbb{R}^3}\frac{e^{i(\xi x + \eta y + \zeta z)}}{2\xi}\frac{\partial}{\partial\bar{k}}\frac{1}{\frac{i\eta-\zeta+\xi^2}{2\xi}+k}\,d\xi d\eta d\zeta.
	\end{equation}
	
	By operating with equation (3.9), we can obtain
	\begin{equation}\label{eq:3.11}
	\frac{\partial}{\partial\bar{k}}\frac{1}{\frac{i\eta-\zeta+\xi^2}{2\xi}+k} = \pi \delta(\frac{\xi^2-\zeta}{2\xi}+k_R) \delta(\frac{\eta}{2\xi}+k_I).
	\end{equation}
	
	Substitute equation (3.11) back into the right side of equation (3.10), we obtain
	\begin{equation}\label{eq:3.12}
		\frac{\partial G}{\partial \bar{k}} = -\frac{1}{4\pi^2} \int_{\mathbb{R}^3} \xi
		e^{i \left[\xi x -2\xi k_I y + \left(\xi^2+2\xi k_R\right) z\right] }
		d\xi.
	\end{equation}
	
	Substituting equation (3.12) back into the right-hand side of equation (3.8), obviously we obtain
	\begin{equation}\label{eq:3.13}
		\chi(x, y, z, k) = \int_{\mathbb{R}} \xi \widehat{q}(\xi, k) 
		e^{i\left[\xi x - 2k_I \xi y + (\xi^2 + 2\xi k_R) z\right]} 
		d\xi,
	\end{equation}
	with 
	$$
	\widehat{q}(\xi, k) = -\frac{1}{4\pi^2} \int_{\mathbb{R}^3} 
		e^{-i\left[\xi x' - 2k_I \xi y' + (\xi^2 + 2\xi k_R) z'\right]}\left[\partial_x^{-1}\left(q_y+iq_z\right)-q_x-q^2\right]\mu(x',y',z',k)\,dx'dy'dz'.
	$$	
		
	Now replacing $k$ with $k+\xi$ and applying $e^{i\left[\xi x - 2k_I \xi y + (\xi^2 + 2\xi k_R) z\right]} $ to equation (3.4), that is
	\begin{equation}\label{eq:3.14}
		\begin{small}
		\begin{aligned}
			&
			e^{i\left[\xi x - 2k_I \xi y + (\xi^2 + 2\xi k_R) z\right]}  \mu(x,y,z,k+\xi) = e^{i\left[\xi x - 2k_I \xi y + (\xi^2 + 2\xi k_R) z\right]}  + 
			\\
			&
			\int_{\mathbb{R}^3}\widehat{G}(x-x',y-y',z-z',k,\xi)
			\left[\partial_x^{-1}\left(q_y+iq_z\right)-q_x-q^2\right]
			\mu(x',y',z',k+\xi)dx'dy'dz',
		\end{aligned}
	    \end{small}
	\end{equation}
	where
	$$
	\widehat{G}(x-x',y-y',z-z',k,\xi) = G(x-x',y-y',z-z',k+\xi) e^{i\left[\xi x - 2k_I \xi y + (\xi^2 + 2\xi k_R) z\right]}.
	$$
	
	The following relationship can be easily achieved:
	\begin{equation}\label{eq:3.15}
		\widehat{G}(x-x',y-y',z-z',k,\xi) = G(x-x',y-y',z-z',k).
	\end{equation}
	
	Simultaneously, using equation (3.7) and equation (3.12)-(3.15), we can find
	\begin{equation}\label{eq:3.16}
		\frac{\partial\mu}{\partial\bar{k}}(x,y,z,k) = \int_{\mathbb{R}} \xi \widehat{q}(\xi, k) 
		e^{i\left[\xi x - 2k_I \xi y + (\xi^2 + 2\xi k_R) z\right]} \mu(x,y,z,k+\xi)
		d\xi.
	\end{equation}
	
	Finally, by integrating equation (3.16), equation (3.2) and (3.3) is obtained. \qed
	
	Expanding $\mu(x,y,z,k)$ at $k$ as follows:
	\begin{equation}\label{eq:3.17}
		\mu(x, y, z, k) = 1 + \frac{\mu_1(x,y,z)}{k} + O\left(\frac{1}{k^2}\right), k \to \infty,
	\end{equation}
	substituting equation (3.17) into equation (3.1), and comparing the coefficients of 1/k, we have 
	\begin{equation}\label{eq:3.18}
		q(x,y,z) = 2i \left(\mu_1\left(x,y,z\right)\right)_x,
	\end{equation}
	Utilizing equation (3.2), then we can obtain
	\begin{equation}\label{eq:3.19}
		q(x,y,z) = \frac{2i}{\pi} \partial_x \int_{\mathbb{R}^3} \xi 
		e^{i\left[\xi x - 2h_I \xi y + (\xi^2 + 2\xi h_R) z\right]} \widehat{q}(\xi, h) \mu(x,y,z,h+\xi)
		d\xi dh_R dh_I.
	\end{equation}
	
	\noindent\textbf{Remark 3.1.} 
	In the linear limit of $q$, as $\mu \to 1$, we obtain the Fourier transform pair
	\begin{equation}\label{eq:3.20}
		\widehat{q}(\xi, k) = -\frac{1}{4\pi^2} \int_{\mathbb{R}^3} 
		e^{-i\left[\xi x - 2k_I \xi y + (\xi^2 + 2\xi k_R) z\right]}\left[\partial_x^{-1}\left(q_y+iq_z\right)-q_x-q^2\right]\,dxdydz,
	\end{equation}
	\begin{equation}\label{eq:3.21}
		q(x,y,z) = \frac{2i}{\pi} \partial_x \int_{\mathbb{R}^3} \xi 
		e^{i\left[\xi x - 2k_I \xi y + (\xi^2 + 2\xi k_R) z\right]} \widehat{q}(\xi, k) d\xi dk_R dk_I.
	\end{equation}
	
	The above proposition states: Assuming $q(x,y,z) \in S(\mathbb{R}^3) $ and the norms of $L_1,L_2$ are sufficiently small, let $\mu(x,y,z,k)$ be the solution to equation (3.2), then the nonlinear Fourier transform of $\widehat{q}(\xi,k)$ is $q$, where $q$ is defined by equation (3.19). If $\mu(x,y,z,k)$ is the solution to equation (3.4), then the inverse of the nonlinear Fourier transform of $q$ is $\widehat{q}(\xi,k)$, where $\widehat{q}(\xi,k)$ is defined by equation (3.3).
	
	\section{$\bar{\partial}$-formalism to the (3+1)-dimensional mKP equation}
	
	\textbf{Proposition 4.1.} 
	\textit{Assuming $q(x,y,z) \in S(\mathbb{R}^3) $ and the norms of $L_1,L_2$ are sufficiently small, let $\mu_0(x,y,z,k)$ be the solution to equation (3.4), with $q$ is substituted by $q_0$, $\widehat{q_0}(\xi,k)$ is defined by equation (3.3). Take $\mu(x,y,z,t,k)$ as the solution to equation (3.2), with }
	\begin{equation}\label{eq:4.1}
	\widehat{q}(\xi,k,t) = \widehat{q_0}(\xi,k)e^{\left(T_R+iT_I\right)t},
	\end{equation}
	\textit{where} 
	\begin{equation}\label{eq:4.2}
	\begin{aligned}
		&
		T_R=-3\xi k_I\left(\xi+2k_R\right),
		\\
		&
		T_I=\xi^3+3\xi^2k_R+3\xi \left(k_R^2-k_I^2\right).
	\end{aligned}
	\end{equation}
	
	\textit{Suppose} $q\left(x,y,z,t\right)$ \textit{is defined by equation (3.19), where} $\widehat{q}\left(\xi,k\right)$ \textit{and} $\mu\left(x,y,z,k\right)$ \textit{respectively substituted by} $\widehat{q}\left(\xi,k,t\right)$ \textit{and} $\mu\left(x,y,z,t,k\right)$. \textit{It follows that} $q\left(x,y,z,t\right)$ \textit{is the solution of (3+1)-dimensional mKP equation}
	\begin{equation}\label{eq:4.3}
		q_{t} + \frac{1}{4} q_{xxx} - \frac{3}{2} q^2 q_x + \frac{3}{4} \partial_x^{-1}(q_{yy} - q_{zz} + 2i q_{yz}) + \frac{3}{4} q_x \partial_x^{-1}(q_y + i q_z) = 0,
	\end{equation}
	\textit{with}
	$$
	q(x,y,z,0) = q_0(x,y,z).
	$$
	
	\noindent\textbf{Proof.} 
	By imposing the time dependence on $ \widehat{q_0}\left(\xi,k\right)$, redefining equation (3.21) according to equation (4.1) yields 
	\begin{equation}\label{eq:4.4}
		q(x,y,z,t) = \frac{2i}{\pi} \partial_x \int_{\mathbb{R}^3} \xi 
		e^{i\left[\xi x - 2k_I \xi y + (\xi^2 + 2\xi k_R) z\right]} \widehat{q}(\xi,k,t) d\xi dk_R dk_I.
	\end{equation}  
	
	Substitute $\widehat{q}(\xi,k,t)$ in equation (4.1) into the linear part of equation (4.3), then equation (4.2) can be obtained.
	
	Similar to equation (3.1), perform the same scaling on the $t$-dependent part of the Lax pair (2.22), we obtain
	\begin{equation}\label{eq:4.5}
		\begin{small}
			\begin{aligned}
				&
				\mu_t - \partial_x^{-1} q_t \mu - \left(q_{xx} - \partial_x^3 - 3ik\partial_x^2 +3k^2 \partial_x + ik^3\right) \mu + \frac{1}{2} q^3 \mu 
				\\
				&
				- \frac{3}{2} \left[q_x  \left(\partial_x + ik\right) + q q_x + q^2 \left(\partial_x + ik\right)\right] \mu
				- \frac{3}{2} \partial_x^{-1} \left(q_y + iq_z \right) \left(q - \partial_x - ik\right)\mu = 0.
			\end{aligned}
		\end{small}
	\end{equation}
	
	In this case, substituting $q(x,y,z)$ with $q(x,y,z,t)$ and $\mu(x,y,z,k)$ with $\mu(x,y,z,t,k)$ in the Lax pair (3.1) and (4.5), yields $q(x,y,z,t)$ as the solution to equation (4.3). \qed
	
	Substitute equation (4.2) into equation (4.1) gives
	\begin{equation}\label{eq:4.6}
		\widehat{q}(\xi,k,t) = \widehat{q_0}(\xi,k)e^{\left\{i\left[\xi^3+3\xi^2k_R+3\xi \left(k_R^2-k_I^2\right) \right]-3 k_I \xi \left(\xi+2k_R\right)\right\} t},
	\end{equation}
	which implies that the integral equation for $\mu(x, y, z, k, t)$ is valid only when
	\begin{equation}\label{eq:4.7}
		k_I\xi \left(\xi+2k_R\right) \geq 0.
	\end{equation}
	
	To ensure the above objective, the sign of $k_I,\xi,\left(\xi + 2k_R\right)$, are selected as shown in the table below.
	
	\begin{table}[h]\label{Table 1}
		\caption{Sign of variables}
		\centering
		\begin{tabular}{c|c|c}
			\hline
			$\quad \xi + 2k_R \quad$ & $\quad \quad \xi \quad \quad$ & $\quad \quad k_I \quad \quad$ \\
			\hline
			{$\leq 0$} & $\leq 0$ & $\geq 0$ \\
			\cline{2-3}
			& $\geq 0$ & $\leq 0$ \\
			\hline
			{$\geq 0$} & $\geq 0$ & $\geq 0$ \\
			\cline{2-3}
			& $\leq 0$ & $\leq 0$ \\
			\hline
		\end{tabular}
	\end{table}
	
	Hence, the double integrable over $d\xi,dk_R$ becomes
	\begin{equation}\label{eq:4.8}
		\begin{aligned}
			&
			\int_{-\infty}^0 dk_R \left( \int_{-\infty}^0 d\xi + \int_{-2k_R}^\infty d\xi \right) + \int_0^\infty dk_R \left( \int_{-\infty}^{-2k_R} d\xi + \int_0^\infty d\xi \right) \\
			&+  \int_{-\infty}^0 dk_R \int_0^{-2k_R} d\xi + \int_0^\infty dk_R \int_{-2k_R}^0 d\xi.
		\end{aligned} 
	\end{equation}

	Introduce the new operator $\digamma_{k_R,k_I,\xi}$ as
	\begin{equation}\label{eq:4.9}
		\begin{aligned}
			\digamma_{k_R, k_I, \xi}[\cdot] &= \int_0^\infty dk_I \left[ \int_{-\infty}^0 dk_R \left( \int_{-\infty}^0 d\xi + \int_{-2k_R}^\infty d\xi \right) + \int_0^\infty dk_R \left( \int_{-\infty}^{-2k_R} d\xi + \int_0^\infty d\xi \right) \right] \\
			&+ \int_{-\infty}^0 dk_I \left( \int_{-\infty}^0 dk_R \int_0^{-2k_R} d\xi + \int_0^\infty dk_R \int_{-2k_R}^0 d\xi \right) [\cdot].
		\end{aligned} 
	\end{equation}
	
	Then we ultimately obtain the expression for the solution of $\mu(x,y,z,t,k)$ as follows:
	\begin{equation}\label{eq:4.10}
		\mu(x,y,z,t,k) = 1 - \frac{1}{\pi} \digamma_{h_R,h_I,\xi} \left[ \Gamma \cdot \frac{\xi \widehat{q_0}(\xi,h) \mu(x,y,z,t,h + \xi)}{h-k} \right],
	\end{equation}
	where
	\begin{equation}\label{eq:4.11}
		\Gamma := \Gamma(x, y, z, t, h_R, h_I, \xi) = \exp[\left(T_R+iT_I\right)t + i\xi x - 2ih_I\xi y + i\xi (\xi + 2h_R)z].
	\end{equation}
	
	Since
	\begin{equation}\label{eq:4.12}
		q(x, y, z, t) = 2i \frac{\partial}{\partial x} \mu_1(x, y, z, t),
	\end{equation}
	with $\mu_1$ corresponding to the $1/k$-term coefficient in the expansion of  $\mu$. According to equation (4.10), that is
	\begin{equation}\label{eq:4.13}
		q(x, y, z, t) = \frac{2i}{\pi} \partial_x \digamma_{h_R,h_I,\xi} [\Gamma \cdot \xi \widehat{q_0} (\xi, h) \mu(x, y, z, t, h+\xi)].
	\end{equation}
	
	\noindent\textbf{Remark 4.1.} 
	In the linear limit of $q$, as $\mu \to 1$, we obtain the Fourier transform pair
	\begin{equation}\label{eq:4.14}
		\widehat{q}(\xi, k) = -\frac{1}{4\pi^2} \int_{\mathbb{R}^3} 
		e^{-i\left[\xi x - 2k_I \xi y + (\xi^2 + 2\xi k_R) z\right]}\left[\partial_x^{-1}\left(q_y+iq_z\right)-q_x-q^2\right]\,dxdydz,
	\end{equation}
	\begin{equation}\label{eq:4.15}
		q(x,y,z) = \frac{2i}{\pi} \partial_x \digamma_{h_R,h_I,\xi} \left[ \xi 
		e^{i\left[\xi x - 2k_I \xi y + (\xi^2 + 2\xi k_R) z\right]} \widehat{q_0}(\xi, k) \right].
	\end{equation} 
	
	Then it can be obtained that
	\begin{equation}\label{eq:4.16}
		q(x,y,z,t) = \frac{2i}{\pi} \partial_x \digamma_{h_R,h_I,\xi} \left[ \Gamma \cdot \xi \widehat{q_0}(\xi,k) \right],
	\end{equation}
	which is a novel $\bar{\partial}$-formalism for the initial value problem of the (3+1)-dimensional mKP equation.
	
	\section{Conclusion}
	In this work, we mainly investigate the integrable extension of the (3+1)-dimensional mKP equation (provided by equation (2.20)) via the $\bar{\partial}$-dressing method. 
	Proposition 3.1. and Proposition 4.1. present the main results of this paper. Specifically, Proposition 3.1. derives the expression for the solution $\mu\left(x,y,z,k\right)$ through spectral analysis and Cauchy-Green formula. While Proposition 4.1., building upon Proposition 3.1., imposes time dependence on $\widehat{q}\left(\xi,k\right)$, thereby providing the $\bar{\partial}$-formalism for the initial value problem of the (3+1)-dimensional mKP equation. 
	
	Interestingly, the extended mKP equation preserves the property of Laplace's equation, that is, the complexification of the mKP equation into a (4+2)-dimensional form can revert to its original (2+1)-dimensional structure by satisfying a particular solution and Laplace's equation. The Fourier transform pair under the linear limit condition of $q$ is given by the Cauchy-Green formula. Similarly, after imposing the time dependence, the Fourier transform pair with the integral operator redefined can also be obtained. Notably, the eigenvalue function and Green’s function play a crucial role in receiving its solution with the  Fourier transform pair. 
	
	\section*{Data Accessibility}
	This article has no additional data.
	
	\section*{Acknowledgments}
	All authors acknowledge support from Beijing Natural Science Foundation (No.1252012) and the National Natural Science Foundation of China (Grant. No.12171475).

	\end{document}